Research Article

# Intrinsic supercurrent diode effect in NbSe$_2$ nanobridge

Yiwen Zhang[1,2], Jiliang Cai[1], Peng Dong[1,2], Jiadian He[1,2], Yifan Ding[1,2], Jinghui Wang[1,2], Xiang Zhou[1,2], Kecheng Cao[1], Yueshen Wu[1,2,*], Jun Li[1,2,†]

[1]*School of Physical Science and Technology, ShanghaiTech University, Shanghai 201210, China*

[2]*ShanghaiTech Laboratory for Topological Physics, ShanghaiTech University, Shanghai 200031, China*

**\* Correspondence to:** Prof. Li, School of Physical Science and Technology, ShanghaiTech University, 393 Middle Huaxia Road, Shanghai 201210, China. Email: lijun3@shanghaitech.edu.cn

Dr. Wu, School of Physical Science and Technology, ShanghaiTech University, 393 Middle Huaxia Road, Shanghai 201210, China. Email: wuysh@shanghaitech.edu.cn

**Abstract**

The significance of the superconducting diode effect lies in its potential application as a fundamental component in the development of next-generation superconducting circuit technology. The stringent operating conditions at low temperatures have posed challenges for the conventional semiconductor diode, primarily due to its exceptionally high resistivity. In response to this limitation, various approaches have emerged to achieve the superconducting diode effect, primarily involving the disruption of inversion symmetry in a two-dimensional superconductor through heterostructure fabrication. In this study, we present a direct observation of the supercurrent diode effect in a NbSe$_2$ nanobridge with a length of approximately 15 nm, created using focused helium ion beam fabrication. Nonreciprocal supercurrents were identified, reaching a peak value of approximately 380 μA for each bias polarity at $B_z^{max} = \pm 0.2$ mT. Notably, the nonreciprocal supercurrent can be toggled by altering the bias polarity. This discovery of the superconducting diode effect introduces a novel avenue and mechanism through nanofabrication on a superconducting flake, offering fresh perspectives for the development of superconducting devices and potential circuits.

## INTRODUCTION

Nonreciprocal charge transport is a phenomenon commonly observed in semiconductors[1,2], characterized by an electron–hole asymmetric junction that produces an asymmetric current in response to positive and negative voltages. This behavior finds extensive applications in electronic devices, including diodes, a.c./d.c. converters, optical isolators, circulators, and microwave diodes across a wide frequency spectrum[3-5]. Among these, the p-n junction, a well-established device in logic and computation, is structured by the hetero-interface of a p-type and an n-type semiconductor, resulting in an asymmetric current-voltage characteristic (IVC). However, the applicability of semiconductor junctions in quantum circuits is limited by the requirement to operate at extremely low temperatures to avoid thermal excitation. To address this challenge, a practical solution is found in the nonreciprocal supercurrent, known as the superconducting diode effect (SDE)[6]. The SDE exhibits nonreciprocity in non-dissipative superconducting current, allowing it to flow exclusively in one direction. Serving as the superconducting counterpart to a semiconducting diode, the SDE has the potential to emerge as a novel non-dissipative circuit element, akin to traditional diodes. This characteristic of SDE opens up exciting possibilities in the realms of superconducting electronics[7], superconducting spintronics[8, 9], and quantum information and communication technology[10, 11].

In recent years, various methods have been proposed to realize the SDE. The initial discovery of SDE was reported by F. Ando et al., who investigated a junction-free superconducting [Nb/V/Ta]$_n$ superlattice, breaking both spatial-inversion and time-reversal symmetries[12]. Subsequent research by the same team introduced another implementation of zero-field SDE using noncentrosymmetric [Nb/V/Co/V/Ta] superconducting films with 20 multilayers, demonstrating the achievability of field-free SDE through noncentrosymmetric superconductor/ferromagnet multilayers[13]. A notable advancement stems from the study of a $NbSe_2/Nb_3Br_8/NbSe_2$ Josephson junction, functioning as a field-free SDE due to the asymmetric Josephson tunneling induced by rotational symmetry breaking from $Nb_3Br_8$ on $NbSe_2/Nb_3Br_8$ interfaces[14]. Moreover, a supercurrent diode effect was observed in few-layer $NbSe_2$ sandwiched between BN. This observation results from the breaking of inversion symmetry caused by the presence of a few layers of $NbSe_2$[15]. The Ising superconductivity nature of the few-layer $NbSe_2$ or rotational symmetry breaking on the $NbSe_2$/BN interfaces may dominate the mechanism. It is crucial to note that prior instances of SDE were based on rotational or time-reversal symmetry breaking on the interface of a two-dimensional superconductor through other quantum material[16-18]. This approach requires a complex technique for interface control, especially concerning electron or magnetic correlation effects on the interface. Therefore, there is a need to explore intrinsic SDE in noncentrosymmetric superconductors. Remarkably, a few-layer $NbSe_2$ has been identified as a noncentrosymmetric superconductor, possessing unique intrinsic Ising-type spin-orbit coupling with a locked electron spin along the out-of-plane axis. Consequently, the pairing symmetry can also be disrupted to generate a nonreciprocal supercurrent[19]. Thus, the observation of intrinsic SDE in low-dimensional patterned $NbSe_2$ appears highly promising.

This study presents the observation of intrinsic Superconducting Diode Effect (SDE) in a $NbSe_2$ nanobridge, created using a focused helium ion microscope. Nonreciprocal critical currents are evident without the need for artificially breaking inversion symmetry when a nonzero magnetic field is applied. *I-V* mapping illustrates the asymmetry of the critical current ($I_c$) concerning the magnetic field under bias polarities at $B_z = \pm 0.2$ mT. A demonstrated example of a bias-polarity-controlled superconductivity diode is provided.

## MATERIALS AND METHODS

Thin NbSe2 flakes are mechanically exfoliated onto a polydimethylsiloxane (PDMS) film positioned on a glass slide. Subsequently, the PDMS is stamped onto the pre-prepared electrode using a micromanipulator located beneath a microscope. The sample transfer process is conducted within a glove box filled with argon gas.

The microbridge sample was fabricated using a Laser Direct-Write Lithography System (Durham Magneto Optics Ltd) and a reactive ion beam etching system (Advance Vacuum Scandinavia AB). **Figure 1**a illustrates the schematic representation of the typical device. The chosen $NbSe_2$ crystal for this study has a thickness of approximately 15.4 nm (shown in **Figure S**1). Electrical contacts and microbridge strips (1 μm wide and 4 μm long) were defined through conventional photolithography, as depicted in **Figure 1**b. The narrow channel serves the purpose of establishing a well-defined current direction in the constriction, indicated by the orange arrow in **Figure 1**a and defined as the $\hat{x}$ direction. The $\hat{z}$ axis is parallel to the stacking direction of $NbSe_2$. Subsequently, the sample was introduced into a

Zeiss Orion NanoFab helium ion microscope, and a 30 kV helium beam was scanned across the microbridges to create nanobridges. A lower linear fluence may slightly influence the superconductivity of NbSe$_2$, whereas a higher linear fluence induces insulating behavior in the barrier. Within these extremes, linear fluences (600 ions/nm in this sample) were determined to selectively diminish superconductivity to a certain depth, effectively reducing the dimension of NbSe$_2$. The regions bombarded by helium ions can be identified by SEM (shown in **Figure 1**c, d).

The as-patterned nanobridge measures approximately 15 nm in length and 1 μm in width, depicted as a red and blue gradient area in **Figure 1**a. The red area signifies the region where superconductivity is disrupted, while the blue area represents NbSe$_2$ reaching the two-dimensional (2D) limit. A schematic diagram of the helium-ion-beam milling process of NbSe$_2$ is presented in **Figure 1**d. During the process, He$^+$ implantation into the NbSe$_2$ crystal can lead to preferential sputtering of selenium atoms within the two-dimensional transition-metal dichalcogenides (TMDs) [20], resulting in the degradation of superconductivity in the irradiated region (red area). Notably, vacancy defects caused by recoil ions serve as the primary source of superconductivity disruption. Consequently, the surface, in contrast to the underlying NbSe$_2$, can be considered nearly undamaged. As the radiation linear fluence increases, the influence depth of He ions extends, subsequently reducing the thickness of the superconducting NbSe$_2$. This process may facilitate the formation of Ising pairing in few-layer NbSe$_2$ (blue area).

Electrical transport measurements were conducted utilizing a Physical Property Measurement System (PPMS-Dynacool, Quantum Design) with the external electric meter comprising a Keithley 2400 as the current source and a Keithley 2182 as the voltage meter. The investigation of transport properties employed conventional four-terminal methods.

## RESULTS AND DISCUSSION

In monolayer or few-layer NbSe$_2$, the in-plane inversion symmetry is disrupted. Consequently, the concurrent impact of the Zeeman effect and substantial intrinsic spin-orbit interactions gives rise to an electron-spin-locking phenomenon along the out-of-plane direction[19,21-23] (**Figure 1**e). Ising-type superconducting pairing symmetry emerges in NbSe2 due to the reverse spin splitting within the valence bands near valleys K and K'. This phenomenon gives rise to intervalley spin-momentum-locked spin-singlet Cooper pairing between two electrons, characterized by opposing momenta and antiparallel out-of-plane spins.

The directions of the inversion symmetry break ($\hat{r}$) and the current ($\hat{I}$) have been illustrated in **Figure 1**e. Consequently, the supercurrent diode behavior of this nanobridge can be influenced by the out-of-plane component ($B_z$) of the magnetic field. When the applied magnetic field ($B$), electric current ($I$), and polar axis ($r$) are mutually orthogonal, the critical current ($I_c$) magnitude depends on both $B$ and $r$, leading to a magnetic chiral effect[24]. Therefore, the supercurrent diode behavior is determined by the out-of-plane component $B_z$.

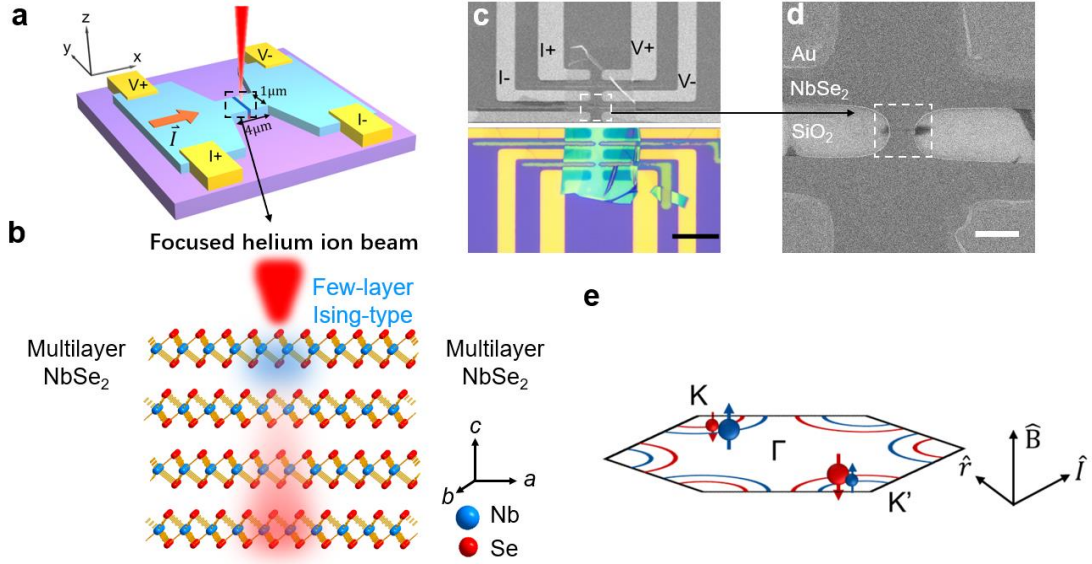

**Figure 1.** a: Schematic illustration of the measurement configuration. The orange arrow indicates the supercurrent pathway, and the red laser indicates the focused helium ion beam creating a weakly coupled junction in the multilayer NbSe$_2$. The central constriction is about 1 μm wide and 15 nm long. The direction of the supercurrent is defined along the x-axis, while the z-axis is perpendicular to the crystal plane. b: Artistic representation of the focused helium ion beam etching the multilayer NbSe$_2$. The crystal structure is that of NbSe$_2$. c: SEM (upper) and optical (lower) image of the device. The nanobridge is made by ion beam etching, and an Ohmic contact is made by the pick-up and transfer method; scale bar, 20 $\mu m$. d: A zoom-in image of the dashed rectangular area shown in b. The color change reflects the change in electrical conductivity, which can be observed. e: Illustration depicting type-I Ising superconductivity: the pairing of electrons in valleys with opposite spin splitting. $\hat{r}$ ($\hat{y}$) refers to the direction of space inversion symmetry breaking, $\hat{B}$ ($\hat{z}$) refers to the direction of time inversion symmetry breaking, and $\hat{I}$ ($\hat{x}$) refers to the current direction.

**Figure 2** illustrates the nonreciprocal transport properties of the NbSe$_2$ nanobridge under varying irradiation linear fluences. **Figures 2** a-c depict three pairs of current-voltage (*I-V*) characteristics corresponding to the magnetic fields applied along $\hat{z}$-axis. *I-V* curves are recorded under $B_z = 0$ (top), $B_z = 0.7$ mT (middle), $B_z = -0.7$ mT (bottom) for three linear fluences. All curves represent zero-to-finite (either positive or negative) bias sweep directions, mitigating potential heating effects. The orange (blue) curve denotes the current density in the nanobridge oriented towards the positive (negative) $\hat{x}$ direction. Two additional devices, fabricated with NbSe$_2$ of identical thickness, were irradiated at different linear fluences. **Figure 2**a and **2**b display the *I-V* curves for the devices under irradiation linear fluences of 0 and 300 ions/nm. Notably, nonreciprocal charge transport is absent in these samples. Given the intrinsic thickness of the NbSe$_2$ flake at 15.4 nm, the Ising Spin-Orbit Coupling (SOC) is likely ineffective, and it is reasonable to assume the preservation of inversion symmetry[22]. The minimal irradiation of 300 ions/nm appears insufficient to impact the superconductivity of the nanobridge region, as evidenced by the absence of nonreciprocal transport.

As the irradiation linear fluence increases to 600 ions/nm, a robust SDE becomes evident, as depicted in the *I-V* curves in **Figure 2**c. Notably, a significant disparity exists in linear fluences between $I_c^+$ and $|I_c^-|$ in the critical current for the two supercurrent orientations. The sign of $\Delta = I_c^+ - |I_c^-|$ changes with the reversal of the magnetic field $B_z$, confirming that Δ is intrinsically determined by the magnetic field. The *I-V* loop further excludes the influence of thermal effects (shown in **Figure S2**). It is essential to note that the nanobridge appears to be slightly damaged under helium ion irradiation, as evidenced by optical microscope and scanning electron microscope images (**Figure 1**c, d). A zoom-in image can clearly show the area of helium ion radiation, framed by a dashed rectangle. We hypothesize that helium ions might influence chemical bonds or even induce doping[25,26], rather than causing a direct etching effect. Similar to the outcomes of electron beam irradiation[27], helium ion irradiation can introduce disorders into thin crystals, potentially suppressing superconductivity. Temperature dependence of resistance (**Figures 2** d-f) exhibits a broadened superconducting transition region with increasing linear fluences, which indicates the degradation of superconductivity in nanobridge. Consequently, our focus shifts to the study of the sample under 600 ions/nm irradiation. The results of another two devices (Device#4 and Device#5) have been shown in **Figure S**3-5.

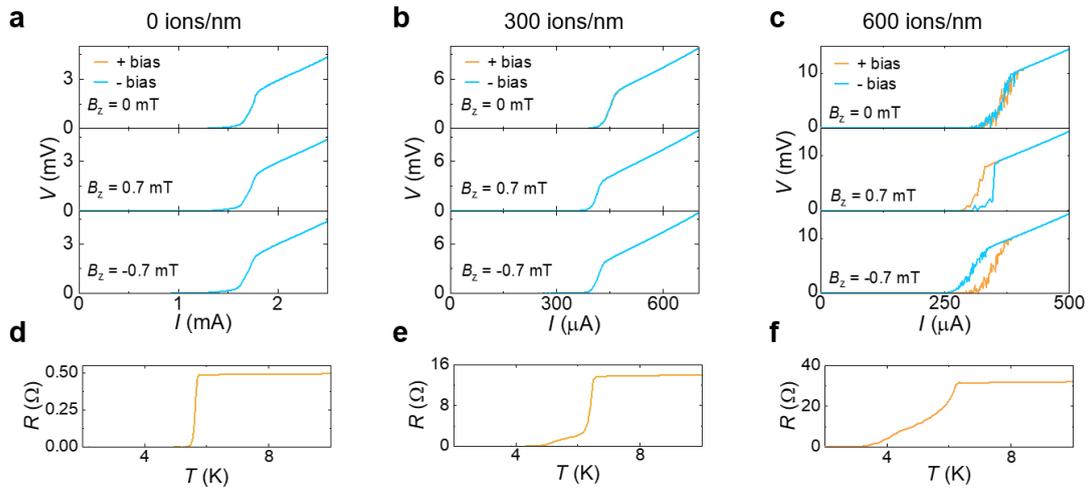

**Figure 2.** a-c: Current-voltage characteristics were measured at 2 K, considering opposite bias polarities (current directions) in a 4-terminal configuration, under zero magnetic field $B_z = 0$ T, which is always applied parallel to z direction. The bias sweep direction consistently proceeds from 0 to a finite bias. But for $B_z = -0.7$ mT. a and b: Measurement for 0 ions/nm and 300 ions/nm sample. c: 600 ions/nm sample. There is a difference between two critical currents. With the orientation of the magnetic field reversed, the roles of the two bias polarities are also exchanged. d-f: The temperature-dependent resistance under different linear fluences ranging from 0 ions/nm to 600 ions/nm.

The *I-V* values at 2 K under various magnetic fields are presented in **Figure 3**a, where the color bar reflects the differential resistance (d*V*/d*I* ). An enlarged view around zero field is provided in **Figure 3**b, revealing a distinct asymmetric behavior. The maximum of the differential resistance is defined as the critical current (*I*c). The *I*c values for both positive ($I_c^+$) and negative ($|I_c^-|$) sides are plotted under different magnetic fields (**Figure 3**c). The range between the positive critical current ($I_c^+$) and the absolute value of the negative critical current ($|I_c^-|$) defines the supercurrent diode regime. In this regime, the current flows without dissipation in only one direction, and this direction can be chosen by altering the sign of the magnetic field. These curves clearly demonstrate that the nonreciprocal components' sign in Ic is exclusively determined by the relative orientation of the current and magnetic field. Regardless of the bias polarity, the critical current increases with the magnetic field, peaking at a nonzero field ($B_{zmax} \approx 0.2$ mT).

The substantial rise in the critical current is vital as it dismisses the possibility of the nonreciprocal supercurrent originating from Joule heating[15]. Although the origin of the supercurrent diode effect in NbSe$_2$ requires further investigation, several potential explanations can be considered. These include the interplay between Meissner currents and barriers for vortex entry[28,29], vortex flow in asymmetric pinning potentials[19], and the influence of valley-Zeeman spin-orbit interaction[15]. It is noteworthy that the observed behavior closely resembles the characteristics of NbSe$_2$ in the 2D limit[19,21,22,30,31], despite our sample having a thickness of approximately 15.4 nm. However, upon dimension reduction in the nanobridge, phenomena akin to those observed in Josephson junctions emerge (refer to **Figure S**4). It is noteworthy that in non-uniform Josephson junctions, the dominance of Josephson vortex motion leads to an asymmetric Fraunhofer pattern[32,33]. A recent experimental study has successfully achieved SDE by introducing non-uniform current and a single Abrikosov vortex in the junction electrodes[34]. Actually, the non-uniform penetration of helium ions can induce non-uniform supercurrent, even when the current is uniform based on the device's geometry, as shown in **Figure S**6. This inhomogeneity can amplify the SDE, particularly in the presence of a large critical current density, 26.7 $mA \cdot \mu m^{-2}$ in this instance.

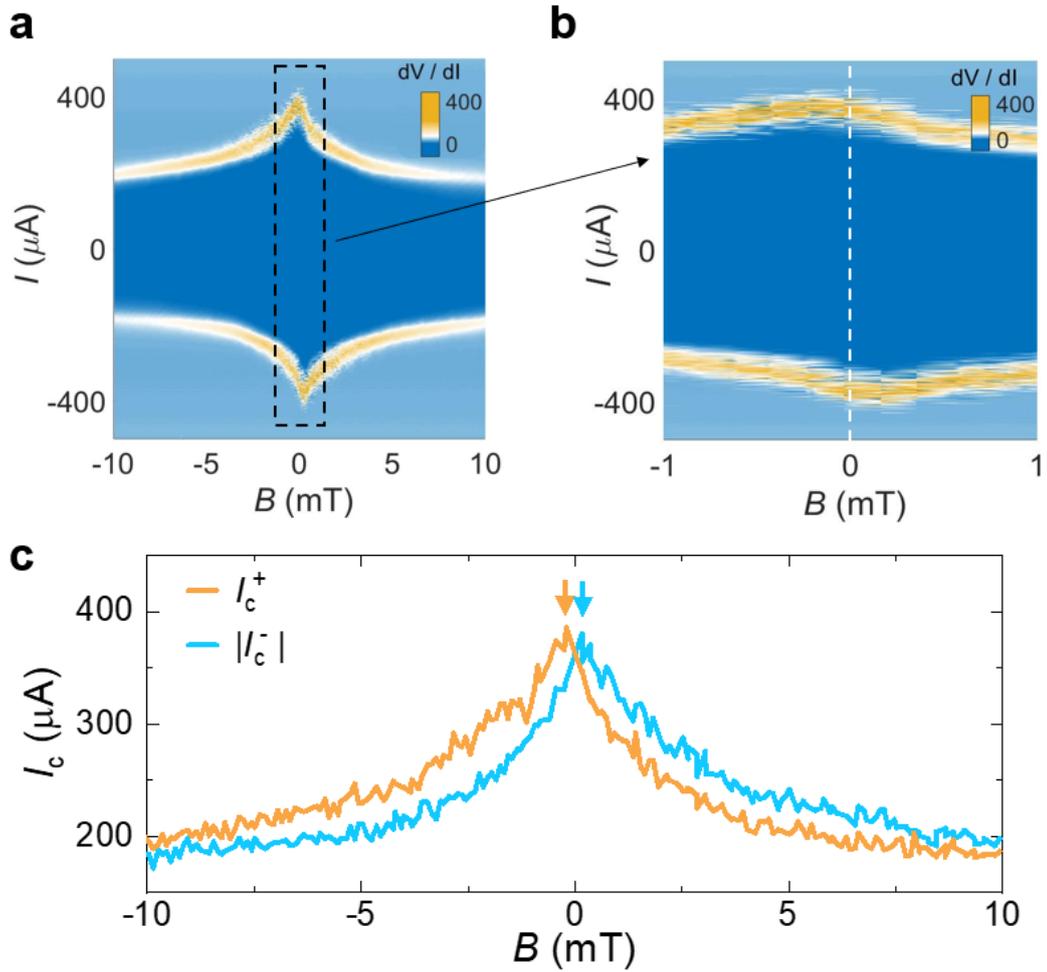

**Figure 3.** a: A color map of critical current in the plane of magnetic field at 2 K, and a color bar shows the value of d$V$/d$I$. An apparent current asymmetry with respect to the magnetic field can be observed. The critical current can be extracted as shown in c. b: A magnification of a. The red dotted line marks the zero-field position. c: Nonreciprocal critical current ($I_c$) as a function of $B_z$, with positive bias (orange) and negative bias (red), exhibiting symmetry across reflection about $B_z$= 0. The critical current peaks at a nonzero |$B_{zmax}$| = 0.2 mT, indicated by labeled orange and blue arrows.

The primary outcome of our observations indicates that the supercurrent diode effect in the NbSe2 nanobridge is governed by the out-of-plane magnetic field. Based on this idea, we demonstrated bias-polarity-controlled SDE at 2 K, 0.2 mT, where the $|I_c^-|$ reaches maximum, as shown in **Figure 4**. The values of $I_c^+$ and $|I_c^-|$ are 345 μA and 380 μA, respectively. A square-wave excitation with an amplitude of 340 μA is applied (**Figure 4**, top panel). The bottom panel of **Figure 4** reveals that the nanobridge maintains the superconducting state with negative current (blue area) and transitions to the normal state during positive current (white area). This outcome strongly suggests that the switch between superconducting and normal conducting states is contingent on the magnetic field's sign and the current direction. Consequently, the nonreciprocity can be readily manipulated by altering the bias polarity of the applied current under a small magnetic field.

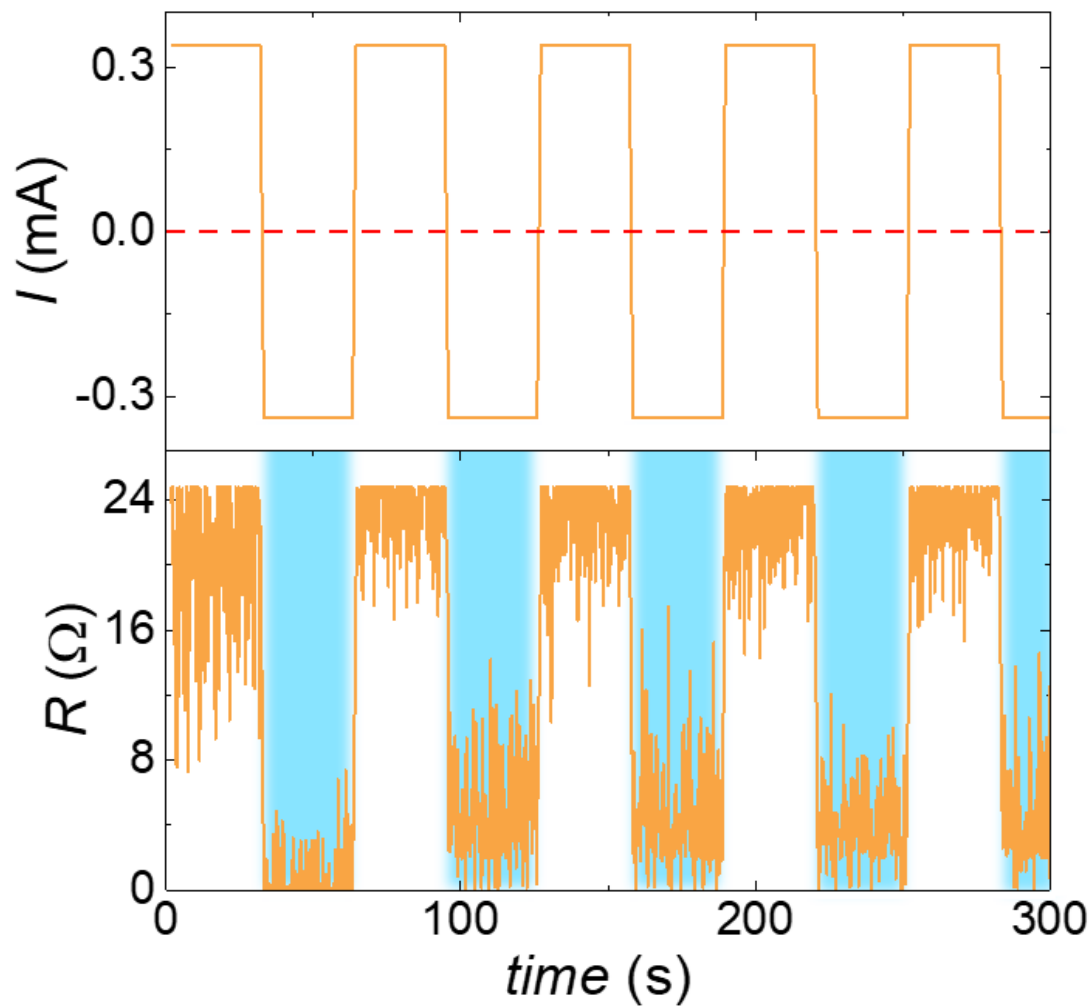

**Figure 4.** Controllable superconducting diode. The top panel displays the square-wave excitation applied at 2 K with an amplitude of 340 μA (between $I_c^+$ and $|I_c^-|$) under a magnetic field of 0.2 mT. The coincidentally measured junction voltage is presented in the bottom panel. In this depiction, the blue shaded region represents the superconducting state, where the voltage remains zero under negative current bias. Conversely, the white region signifies the normal state, with a high voltage observed during positive current bias. The red dotted line denotes the zero line.

## CONCLUSIONS

In conclusion, we have illustrated a supercurrent diode effect in a NbSe$_2$ nanobridge fabricated using a focused helium ion microscope. Our findings indicate that this effect is governed by the out-of-plane magnetic field, presenting a deviation from observations in Rashba superconductors. Nonreciprocal critical currents can be well turned under bias polarities at $B_{zmax} = \pm 0.2$ mT. Helium ions break the superconductivity at a certain depth and shrink the thickness of the superconductor to 2D limit. The findings suggest that the supercurrent diode effect may probably be linked to the inversion symmetry breaking caused by Ising spin-orbit coupling in a few layers of NbSe2 and can be improved by heterogeneous penetration of helium ions. This insight offers a potential avenue for comprehending and optimizing the performance of the supercurrent diode effect, with implications for its application in superconducting logic and memory devices.

## DECLARATIONS

### Acknowledgments

Thanks to the technique support from Soft Matter Nanofabrication Laboratory and The Electron Microscopy Center in the School of Science and Technology in ShanghaiTech University.

### Authors' contributions

Made substantial contributions to conception and design of the study and performed data analysis and interpretation: Yiwen Zhang.

Transmission electron microscopy (TEM) and ion damage calculation: Jiliang Cai, Kecheng Cao.

Writing, review and editing: Yueshen Wu, Jun Li.

Investigation: Peng Dong, Jiadian He, Yifan Ding, Jinghui Wang, Xiang Zhou

### Availability of data and materials

Not applicable

### Financial support and sponsorship

This research was supported in part by the Ministry of Science and Technology (MOST) of China (No. 2022YFA1603903), the National Natural Science Foundation of China (Grants No. 12004251, 12104302, 12104303), the Science and Technology Commission of Shanghai Municipality, the Shanghai Sailing Program (Grant No. 21YF1429200), the start-up funding from ShanghaiTech University, Beijing National Laboratory for Condensed Matter Physics, the Interdisciplinary Program of Wuhan National High Magnetic Field Center (WHMFC202124).

### Conflicts of interest

All authors declared that there are no conflicts of interest.

**Copyright**

© The Author(s) 2023.